\documentclass{ws-procs9x6}

\begin{document}

\title{HOW TO USE MOLECULAR CLOUDS TO STUDY THE PROPAGATION OF COSMIC RAYS IN THE GALAXY}

\author{S. GABICI$^*$}

\address{Astroparticule et Cosmologie (APC), CNRS, UMR7164, Paris, France\\
$^*$E-mail: stefano.gabici@apc.univ-paris7.fr}

\begin{abstract}
Observations of molecular clouds in the gamma ray domain provide us with a tool to study the distribution of cosmic rays in the Galaxy.
This is because cosmic rays can penetrate molecular clouds, undergo hadronic interactions in the dense gas, and produce neutral pions that in turn decay into gamma rays. The detection of this radiation allows us to estimate the spectrum and intensity of cosmic rays at the cloud's position. Remarkably, this fact can be used to constrain the cosmic ray diffusion coefficient at specific locations in the Galaxy.
\end{abstract}


\bodymatter


\vspace{1cm}

\noindent
Galactic Cosmic Rays (CRs) are believed to be accelerated by SuperNova Remnants (SNRs).\cite{hillas} The main (but not conclusive) arguments supporting this idea are the fact that SNRs can provide the total power required to maintain the galactic CR population, and the belief that an efficient acceleration mechanism, diffusive shock acceleration \cite{bell}, operates in these objects. The detection of some SNRs in TeV gamma rays \cite{felixreview}, expected if they indeed are the sources of CRs \cite{luke}, also supports this scenario, though the emission might have a leptonic origin and not be related to the acceleration of CRs. \cite{don}  

Besides the total CR power of the Galaxy, a theory of CR origin must also explain the spectrum, isotropy, and chemical composition of CRs.\cite{book}~This requires a thorough knowledge not only of the nature of CR sources, but also of the way in which CRs propagate in the Galaxy. This is because the observed properties of CRs results from the equilibrium between the injection rate of CRs from sources and their diffusive escape from the Galaxy.

Measurements of the amount of spallation suffered by CRs allow us~to infer the average residence time of a CR of energy $E_p$ in the Galaxy as $t_{res} \propto E_p^{-\delta}$, with $\delta \sim 0.3 - 0.7$. If $h$ is the length a CR has to move away from its source before escaping the Galaxy (i.e. the Galaxy's thickness), then the diffusion coefficient reads: $D_{gal} \approx h^2/t_{res} \approx 10^{28} (E_p/10~{\rm GeV})^{\delta}~{\rm cm^2/s}$. 
However, this has to be intended as the average diffusion coefficient in the Galaxy, and local variations (both in time and space) might exist.

In particular, the diffusion coefficient might be suppressed close to CR sources. This is because CRs can excite magnetic turbulence while streaming away from their acceleration site. This would enhance the scattering rate of CR themselves and thus reduce the diffusion coefficient \cite{kulsrud}. The problem of estimating, on theoretical grounds, the diffusion coefficient around CR sources is far from being solved, mainly because of its intrinsic non-linearity and because various mechanisms might damp the CR--generated waves and thus affect the way in which CRs diffuse  \cite{farmer,ptuskin}. 

In principle, gamma ray observations can provide us with constraints~on the diffusion coefficient close to CR sources. Once escaped from their sources, CRs undergo hadronic interactions with the surrounding gas and produce gamma rays. The characteristics of such radiation (in particular~its spectrum and intensity as a function of the time elapsed since CRs escaped the source) depend on the value of the diffusion coefficient that can thus~be constrained, if a reliable model for CR acceleration at the source is available. The presence of massive Molecular Clouds (MCs) close to the source would enhance the gamma ray emission, making its detection more probable.
Studying such radiation is of great importance not only in order to reach a better understanding of how CRs diffuse, but also because its detection can provide an indirect way to identify the sources of galactic CRs \cite{montmerle,casse,atoyan,GA2007,GAC2009}.

\section*{Molecular clouds as cosmic ray barometers}

Consider a MC with mass $M_{cl}$ at a distance $d$, located in a region in the Galaxy where the CR (protons) intensity is $J_{CR}$. 
For illustrative purpose, we set here $J_{CR} = K_{CR} E_p^{-\alpha}$. We further assume that the high energy CRs (the ones with energy above the threshold for $\pi^0$-production) can freely penetrate the cloud\cite{confinement}
. Under these assumptions the expected gamma ray flux from the MC due to proton-proton interactions is given by:
\begin{equation}
\label{flux}
F_{\gamma}(E_{\gamma}) \sim Y_{\gamma} ~ \frac{\sigma_{pp}}{m_p} ~ J_{CR}(E_{\gamma}) ~ \left( \frac{M_{cl}}{d^2} \right) \propto E_{\gamma}^{-\alpha}
\end{equation}
where $\sigma_{pp} \approx 34 ~ {\rm mb}$ is the interaction cross section
, $m_p$ is the proton mass, and $Y_{\gamma}$ depends on $\alpha$ and is tabulated in Ref.~\refcite{book}. Equation~\ref{flux} is valid at high energies only ($E_{\gamma} \gtrsim 10~{\rm GeV}$), while at lower energies the spectrum (in log-log scale) is symmetric with respect to the energy $E_{\gamma} = m_{\pi^0}/2 \sim 70~{\rm MeV}$. 

Assume now that the CR intensity in the region under exam differs by a factor $\delta(E_p) = J_{CR}(E_p)/J_{bg}(E_p)$ from the one measured at the Earth (which is:\cite{pdg} $(4 \pi/c) E_p^2 J_{bg}(E_p) \sim 6 \times 10^{-3} (E_p/{\rm TeV})^{-0.7} {\rm eV~cm^{-3}}$), so that Equation~\ref{flux} can be rewritten as:
\begin{equation}
\label{barometer}
E_{\gamma}^2F_{\gamma}(E_{\gamma}) \sim 2.5 \times 10^{-13} \left( \frac{Y_{\gamma} f_{0.1}^{2.7-\alpha}}{0.0275} \right) \left( \frac{M_5}{d_{kpc}^2} \right) \delta(E_p) E_{\gamma}^{-0.7} {\rm TeV/cm^2/s}
\end{equation}
where $M_5$ is the mass of the MC in units of $10^5 M_{\odot}$, $d_{kpc}$ is the distance in kpc, and $f_{0.1} = (f/0.1) \approx 1$ takes into account the fact that on average CRs with energy $E_p$ produce gamma rays with energy $E_{\gamma} \approx f \times E_p$. Moreover, a multiplicative factor of 1.5 has been applied to account for the contribution to the emission from nuclei heavier than H both in CRs and in the MC gas.

If gamma rays are detected from a MC, and its mass and distance are known from its CO emission\cite{dame}, then Equation~\ref{barometer} allows one to measure both~the spectrum and intensity of CRs at the MC's location, and thus the quantity $\delta(E_p)$. It follows that MCs can be effectively used as probes of the energy density of CRs at different locations in the Galaxy, and for this reason have been sometimes referred to as {\it CR barometers}. \cite{issa,felixclouds,sabrina1} Since observations in the~GeV range of the Galaxy suggest that, on large spatial scales, CR variations are not very large, \cite{cosB,egret,luigi} $\delta(E_p)$ is normally interpreted as the excess above the galactic CR background. Some examples of how to use this fact to estimate the CR diffusion coefficient are given in the following sections.



\section*{The galactic centre ridge}

The H.E.S.S. collaboration reported on the detection of diffuse gamma ray emission from the Galactic Centre (GC) ridge. \cite{ridge} Remarkably, the emission correlates spatially with a complex of giant MCs, suggesting that the emission itself is likely to be originated by CR hadronic interactions in the dense ambient gas. The TeV gamma ray spectrum extracted from the region of galactic coordinates $|l| < 0.8^{\circ}$, $|b| < 0.3^{\circ}$ can be fitted by a power law with index $\alpha \sim 2.29$ and normalization $\sim 1.73 \times 10^{-8} {\rm TeV^{-1} cm^{-2} s^{-1} sr^{-1}}$. At the distance of the GC ($\sim 8.5~{\rm kpc}$) the emitting region has a size of $\approx 240 \times 90 ~ {\rm pc}$, and encloses a total gas mass of $1.7 - 4.4 \times 10^7 M_{\odot}$. \cite{ridge}

By assuming that all the observed gamma ray emission has an hadronic origin, we can use Equation~\ref{barometer} to show that, close to the GC, the CR intensity at 10 TeV is enhanced with respect to the local one by a factor of $\delta \approx 3 - 9$, and grows at higher energies due to the hardness of the measured emission. Moreover, the total energy in form of CRs needed to explain~the emission is $\approx 10^{50} {\rm erg}$, if the observed spectrum is extrapolated from 1~GeV to 1~PeV. Thus, a single SNR with explosion energy $E_{SN} \approx 10^{51} {\rm erg}$ and acceleration efficiency $\eta \approx 10\%$ might explain the whole CR excess in the region. \cite{ridge}

One possible CR accelerator is the SNR Sgr A East, located close to the GC and with an estimated (but uncertain) age of $t_{age} \approx 10~{\rm kyr}$. If~the CRs responsible for the observed emission are assumed to be released during~the earliest phase of the SNR evolution, then it is possible to estimate the diffusion coefficient by requiring them to fill the emitting region of size $l_{\gamma} \approx 120~{\rm pc}$ in a time $t_{age}$. This gives $D \approx l_{\gamma}^2/4 ~ t_{age} \approx 10^{29} {\rm cm^2/s}$, which, taking into account the large uncertainties in the assumptions made, is in rough agreement with (maybe somewhat smaller than \cite{ridge}) the value of the diffusion coefficient of $\sim$ TeV CRs in the galactic disk. The presence of a massive MC at $l = 1.3^{\circ}$ ($\sim 200 ~ {\rm pc}$) which does not show prominent gamma ray emission suggests that CRs accelerated at the GC have not diffused yet up to that distance, and this justifies our assumption for $l_{\gamma}$. \cite{ridge}

This estimate of the diffusion coefficient has to be regarded as a very rough one, due to the presence of many uncertainties in its derivation. For example, another source (e.g. the black hole Sgr A$^*$ \cite{melia}), or many sources (e.g. an ensemble of SNRs \cite{wolfendale}) might be the accelerators of the CRs responsible for the TeV emission. Moreover, proton energy losses (ignored here) might affect the estimate of the diffusion coefficient by a factor of $\approx 2$. \cite{apostolos}

Better constraints on the diffusion coefficient and on the nature of the source(s) of the accelerated particles could be obtained if the spatial variation of the spectrum of the gamma ray emission was known. Measuring such variation, though challenging for currently operating instruments, is within the capabilities of future ones like the Cherenkov Telescope Array. It seems thus appropriate to discuss the future perspectives in this direction.

According to the scenario summarized above, the SNR Sgr~A East, located close to the GC, released about $10^{50}~{\rm erg}$ in form of CRs about 10~kyr ago. Since high energy CRs are believed to be accelerated by young SNRs (aged up to few thousands years), we are allowed to consider this as an impulsive injection event. If CRs are injected with a power law spectrum $\propto E^{-2}$, their time dependent spatial distribution around the source is: \cite{atoyan}
\begin{equation}
\label{atoyan}
N_{CR}(R,E_p,t) = \frac{\eta ~ E_{SN}}{\pi^{3/2} \ln(E_p^{max}/E_p^{min})} ~ \frac{E_p^{-2}}{R_d^3} ~ e^{-\left(\frac{R}{R_d}\right)^2}
\end{equation}
where 
$E_{p,max}$($E_{p,min}$) is the maximum (minimum) CR energy, $R_d=\sqrt{4\,D(E_p)\,t}$ is a diffusion length and $D$ the diffusion coefficient. The CR intensity is the sum of Equation~\ref{atoyan} and the galactic CR background.

\begin{figure}[t]
 \centering
 \includegraphics[width=0.5\textwidth]{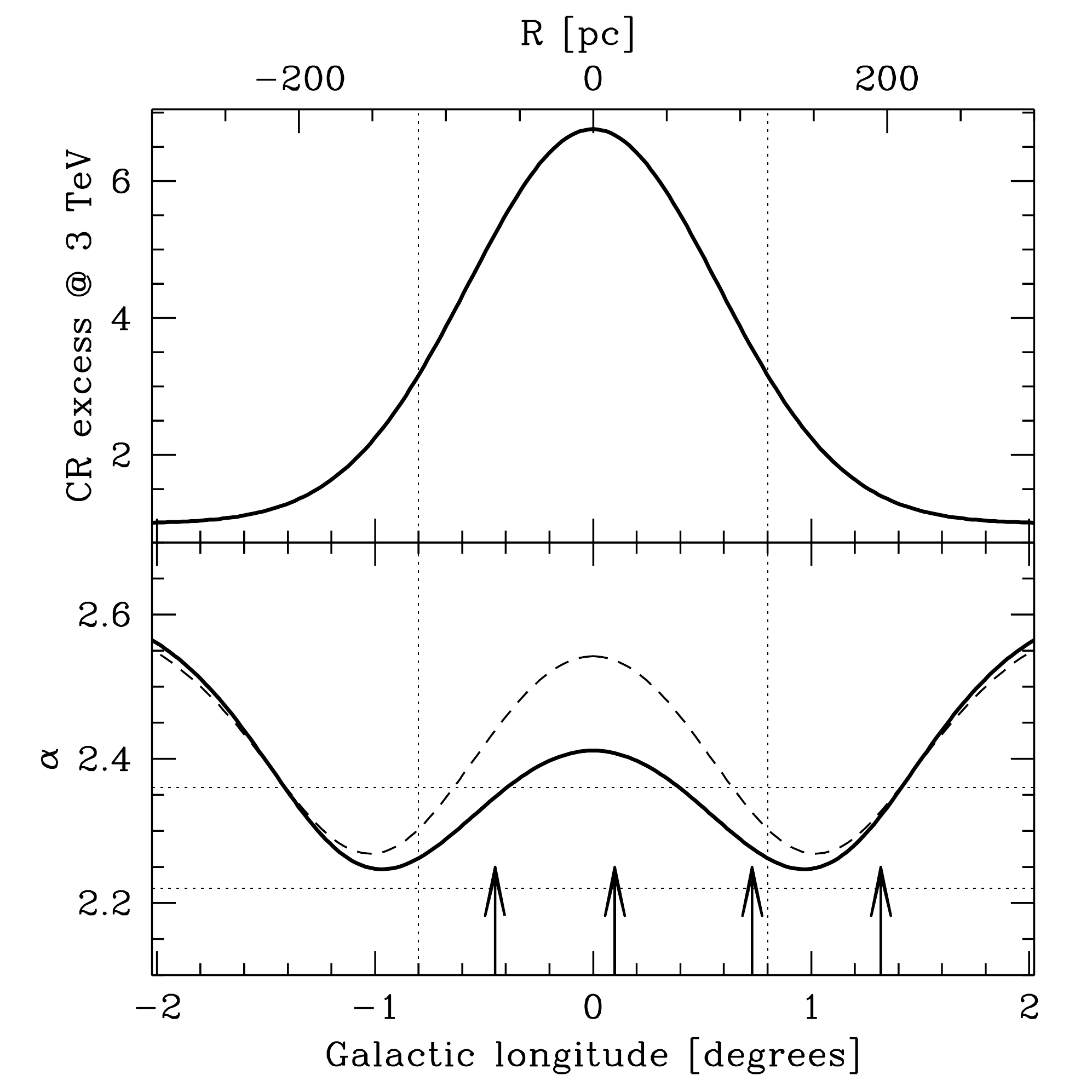}      
  \caption{Upper panel: excess of 3 TeV CRs over the CR background. Lower panel: slope of the gamma ray emissivity at 300 GeV for $\delta = 0.3$ (solid) and $0.4$ (dashed). Horizontal lines represent the observed slope (measured value $\pm$ error). Dotted vertical lines delimit the emitting region. Arrows refer to the positions of the peaks in the gas distribution.}
  \label{fig:excess}
\end{figure}

The expected excess over the background for CRs with energy 3 TeV~is shown in Figure~\ref{fig:excess} (top panel) as a function of galactic longitude. These are the CRs that produce most of the emission detected above $\sim300\,{\rm GeV}$ in the region $|l| < 0.8^{\circ}$ (marked by dotted vertical lines in Figure~\ref{fig:excess}). The slope of the gamma ray emissivity at 300 GeV and for a diffusion coefficient $D \propto E^{\delta}$ with $\delta = 0.3$ is plotted as a solid line in the bottom panel, where it is compared with the measured one, represented by the horizontal dotted lines (measured slope $\pm$ statistical error). 
The arrows represent the (projected) position of the four prominent peaks in the gas density distribution. If the total gas mass is concentrated at these locations and if projection effects in the gas distribution do not play an important role, the spectral slope of the emissivity should roughly represent the observed slope also.

The results in Figure~\ref{fig:excess} can be interpreted as follows. Within roughly one diffusion length, the gamma ray spectrum hardens with the distance from the source. This is because higher energy CRs occupy a larger volume of space around the source, and thus the low energy emission is progressively suppressed at larger and larger distances. On the other hand, at distances much larger than one diffusion length, the CR excess vanishes and the steep spectrum characteristic of the CR background is recovered. Though these calculations are far too crude, they might still serve to estimate the order of magnitude of the expected spectral variations within the emitting region. From Figure~\ref{fig:excess} one sees that variations of the order of $\Delta \alpha \approx 0.1 - 0.2$ are expected for a diffusion coefficient with slope $\delta = 0.3$. A stronger dependency on energy would produce more pronounced variations, as illustrated by the dashed line which refers to $\delta = 0.4$. Values of $\delta$ significantly larger than 0.3 would require a harder injection spectrum of CRs in order to preserve the observed slope in gamma rays. 
Such variations in the slope of the emission, if detected, would support the scenario of a localized injection of CRs close to the GC and provide better constraints on the diffusion coefficient.
Complementary studies of the three dimensional distribution of the gas in the region are also needed in order to control projection effects. \cite{3D}

\section*{Molecular clouds in the W28 region}

W28 is a SNR in its radiative phase of evolution, located at a distance~of $\sim 2~{\rm kpc}$, in a region rich of dense molecular gas. 
Gamma ray emission has been detected from the surroundings of W28 both at TeV \cite{hess} and GeV energies \cite{fermi,agile}, by HESS, FERMI, and AGILE, respectively. The TeV emission correlates quite well with the position of three massive ($\approx 10^5 M_{\odot}$) MCs, one of which is interacting with the north-eastern part of the shell (and corresponds to the TeV source HESS J1801-233), and the other two being located to the south of the SNR (TeV sources HESS J1800-240 A and B) . 

Let us assume that this gamma ray emission is the result of hadronic interactions of CRs that have been accelerated at the SNR and then escaped in the surrounding medium.\cite{sf2a} In this scenario, the distribution of runaway CRs around the SNR is still given by Equation~\ref{atoyan}. A diffusion coefficient  $D = \chi D_{gal} \propto E^{0.5}$ is assumed, where $\chi$ represents possible deviations with respect to the average diffusion coefficient in the Galaxy. Equation~\ref{atoyan} tells us that, up to a distance equal to the diffusion radius $R_d$, the spatial distribution of CRs around the source is roughly constant, and given by $N_{CR} \propto \eta E_{SN}/R_d^3$. 
On the other hand, the observed gamma ray flux from each one of the MCs is: $F_{\gamma} \propto N_{CR} M_{cl}/d^2$. 
By using the definitions of $N_{CR}$ and $R_d$ we can finally write the approximate equation, valid within a distance $R_d$ from the SNR:
\begin{equation}
F_{\gamma} \propto \frac{\eta ~ E_{SN}}{(\chi ~ D_{gal} ~ t_{age})^{3/2}} \left( \frac{M_{cl}}{d^2} \right) .
\label{eq:w28}
\end{equation}
Estimates can be obtained from observations for all the physical quantities in Equation~\ref{eq:w28} except for the CR acceleration efficiency $\eta$ and the diffusion coefficient $\chi D_{gal}$. By fitting the TeV data we can thus attempt to constrain 
a combination of these two parameters (namely $\eta/\chi^{3/2}$). 
Given all the uncertainties above, our results have to be interpreted as a proof of concept of the fact that gamma ray observations can serve as tools to estimate the diffusion coefficient. More detections of SNR/MC associations are needed in order to check whether the scenario described here applies to a whole class of objects and not only to a test-case as W28. Future observations from the Cherenkov Telescope Array will most likely solve this issue.

\begin{figure}[t!]
 \centering
 \includegraphics[width=0.5\textwidth]{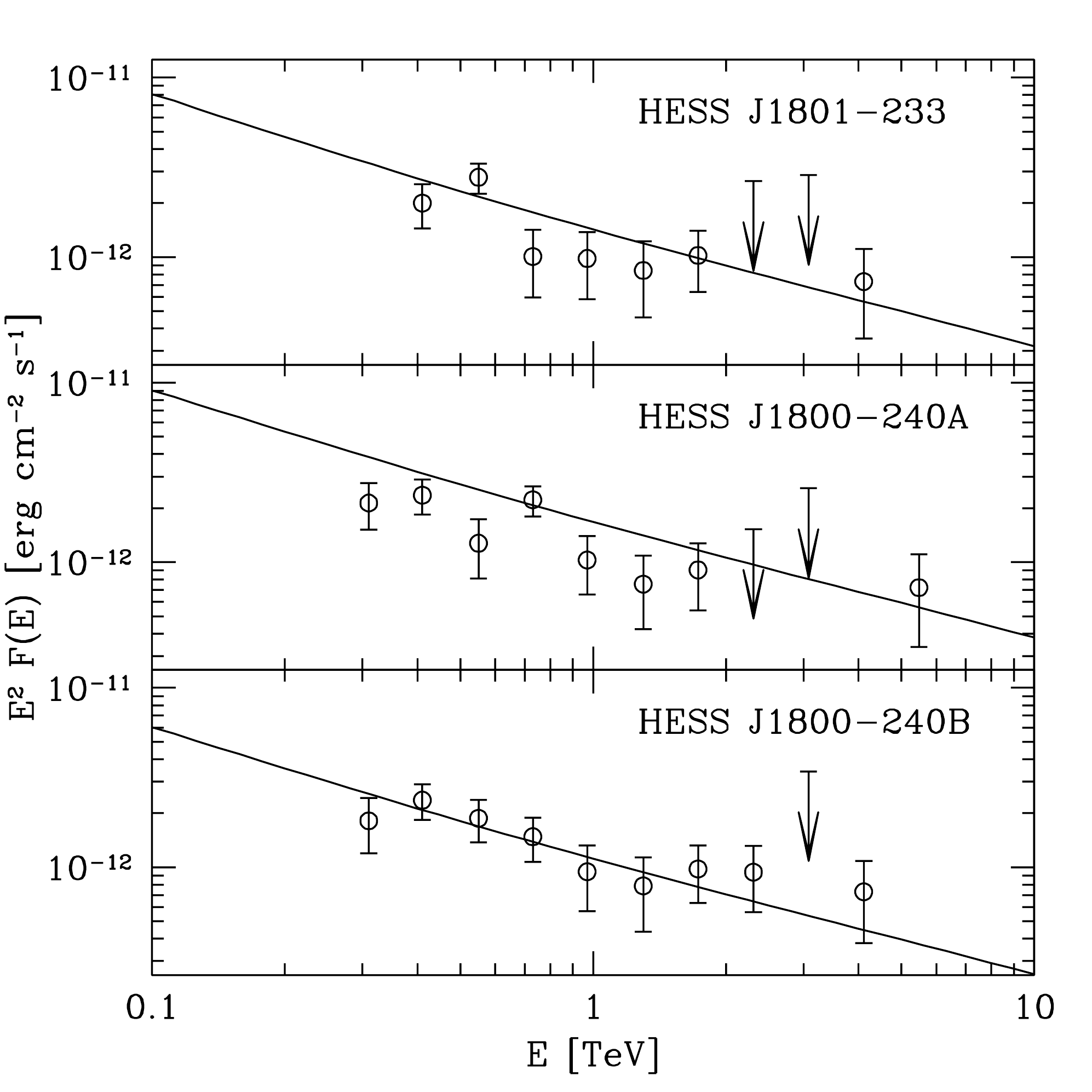}      
  \caption{Simultaneous fit to the three TeV sources detected by HESS in the W28 region. Figure from Ref.~\refcite{sf2a}. See text for more details.}
  \label{fig:w28}
\end{figure}

Figure \ref{fig:w28} shows a simultaneous fit to the HESS data for the three massive MCs in the W28 region, obtained by setting  $\eta/\chi^{3/2} \approx 20$. This implies that the normalized diffusion coefficient $\chi$ has to be much smaller than 1 for any reasonable value of $\eta < 1$. For example, an acceleration efficiency $\eta = 30\%$ corresponds to $\chi = 0.06$, which in turn gives a diffusion distance for TeV particles of $R_d \approx 60~{\rm pc}$. This means that the results in Figure~\ref{fig:w28} are valid if the physical (not projected) distances between the MCs and the SNRs do not significantly exceed $R_d$.  Small values of the diffusion coefficient have been also proposed in Ref.~\refcite{agile,fujita,li}. Note that, since we are considering gamma rays in a quite narrow energy band around $\approx 1$~TeV, we are actually constraining the diffusion coefficient of CRs with energy $\approx 10$~TeV. 
With some caveats \cite{sf2a,ohira}, observations in the GeV range might be used to constrain the diffusion coefficient down to GeV particle energies. 



\section*{IC 443 and other examples}

The SNR IC443 has an estimated age of $t_{age} \approx 3 \times 10^4~{\rm yr}$, explosion energy of $E_{SN} \gtrsim 4 \times 10^{50} {\rm erg}$, and distance $\approx~1.5~{\rm kpc}$.\cite{diego} It has been detected at both GeV \cite{IC443GeV} and TeV \cite{IC443TeV} energies. A power law fit to the TeV emission is: $F_{\gamma} \sim 10^{-11} (E_{\gamma}/0.4\,{\rm TeV})^{-3.1} {\rm cm^{-2} s^{-1} TeV^{-1}}$. Observations of the CO line indicate that about $10^4 M_{\odot}$ of molecular gas are present in the region. \cite{diego} Though the GeV emission might be explained as the emission from shocked clouds,\cite{yas} this scenario cannot account for the very high energy CRs that originate the TeV photons. If these photons are produced by runaway CRs we can attempt to estimate the diffusion coefficient, as done for W28. If the TeV radiation is produced in a MC of mass $\sim 10^4 M_{\odot}$, then from Equation~\ref{barometer} we can infer an excess in the intensty of multi-TeV CRs of the order of $\sim 100$. If these particles had time to diffuse up to a distance $R_d$, their total (integrated over volume) energy is $\pi^{3/2} E_p^2 N_{CR}(0,E_p,t_{age}) R_d^3 \approx 3 \times 10^{47} (R_d/10~{\rm pc})^3~{\rm erg}$ (see Equation~\ref{atoyan}). This number has to be multiplied by $\approx \ln(E_{p}^{max}/E_{p}^{min})$ to account for the extension of the spectrum of CRs injected by the SNR and then equated to $\eta~E_{SN}$. This gives (very roughly) a diffusion coefficient for multi-TeV CRs of $D \approx 10^{27} (\eta/0.1)^{2/3}~{\rm cm^2/s}$. This is, again, much smaller than the average diffusion coefficient in the Galaxy, \cite{diego} and corresponds to $R_d \approx 20~{\rm pc}$. 

Other SNR/MC associations detected in TeV gamma rays that would deserve attention include CTB 37A, HESS J1745-303, and W51. \cite{various} The approach described here might now be extended to the GeV range, where several SNR/MC associations are being detected by FERMI.\cite{slane} The Cygnus region, rich of dense gas and detected at multi-TeV energies, also promises to be an ideal region of the sky for this kind of studies. \cite{cygnus}

\section*{Conclusions}

MCs, when detected in gamma rays, can serve as probes of the CR intensity throughout the Galaxy. If they are located close to CR sources, the intensity of the gamma ray emission might impose constraints on the local diffusion coefficient. In at least one case (SNR W28), preliminary evidence has been reported for a significant suppression of the diffusion coefficient with respect to the average galactic one. This suppression might be the result of an enhancement in the magnetic turbulence due to the streaming of CRs away from the source. Observations with future facilities such as the Cherenkov Telescope Array will conclusively test the feasibility and reliability of the approach to constrain CR diffusion that has been presented here.
 
 \noindent
{\it  Support from the EU is acknowledged [FP7 - grant agr. n$^o$256464].}


\begin{thebibliography}{100}
\bibitem{hillas} A.M. Hillas, {\it J. Phys. G: Nucl. Phys.} {\bf 31}, 95 (2005)
\bibitem{bell} A.R. Bell, {\it Mon. Not. R. Astron. Soc.} {\bf 182}, 147 (1978)
\bibitem{felixreview} F. Aharonian, et al., {\it Rep. Prog. Phys.} {\bf 71}, 096901 (2008)
\bibitem{luke} L.O'C. Drury, et al., {\it Astron. Astrophys.} {\bf 287}, 959 (1994)
\bibitem{don} D.~C. Ellison, et al., {\it Astrophys. J.} {\bf 712}, 287 (2010)
\bibitem{book} V.S. Berezinskii, et al. 1990, {\it Astrophysics of Cosmic Rays} (Amsterdam: North-Holland)
\bibitem{kulsrud} R. Kulsrud, W.P. Pearce, {\it Astrophys. J.} {\bf 156}, 445 (1969)
\bibitem{farmer} A.J. Farmer, P. Goldreich, {\it Astrophys. J.} {\bf 604}, 671 (2004)
\bibitem{ptuskin} V.S. Ptuskin, et al., {\it Adv. Space Res.} {\bf 42}, 486 (2008)
\bibitem{montmerle} T. Montmerle, {\it Astrophys. J.} {\bf 231}, 95 (1979)
\bibitem{casse} M. Cass\'e, J.A. Paul, {\it Astrophys. J.} {\bf 237}, 236 (1980)
\bibitem{atoyan} F.A. Aharonian, A. Atoyan, {\it Astron. Astrophys.} {\bf 309}, 917 (1996)
\bibitem{GA2007} S. Gabici, F.A. Aharonian, {\it Astrophys. J.} {\bf 665}, L131 (2007)
\bibitem{GAC2009} S. Gabici, et al., {\it Mon. Not. R. Astron. Soc.} {\bf 396}, 1629 (2009)
\bibitem{confinement} For a discussion on CR exclusion from MCs see: J. Skilling, A.W. Strong, {\it Astron. Astrophys.} {\bf 53}, 253 (1976); S. Gabici, et al., {\it Astrophys. Space Sci.}~{\bf 309}, 365 (2007); R.J. Protheroe, et al., {\it Mon. Not. R. Astron. Soc.} {\bf 390}, 683 (2008)
\bibitem{pdg} K. Nakamura et al. (Particle Data Group), {\it J. Phys. G} {\bf 37}, 075021 (2010)
\bibitem{dame} T.M. Dame, et al., {\it Astrophys. J.} {\bf 547}, 792 (2001)
\bibitem{issa} M.R. Issa, A.W. Wolfendale, {\it Nature} {\bf 292}, 430 (1981)
\bibitem{felixclouds} F.A. Aharonian, {\it Astrphys. Space Sci.} {\bf 180}, 305 (1991)
\bibitem{sabrina1} S. Casanova, et al., {\it Pub. Astron. Soc. Japan} {\bf 62}, 769 (2010)
\bibitem{cosB} A.W. Strong, et al., {\it Astron. Astrophys.} {\bf 207}, 1 (1988)
\bibitem{egret} S.D. Hunter, et al., {\it Astrophys. J.} {\bf 481}, 205 (1997)
\bibitem{luigi} A.A. Abdo, et al., {\it Astrophys. J.} {\bf 710}, 133 (2010) 
\bibitem{ridge} F.A. Aharonian, et al., {\it Nature} {\bf 439}, 695 (2006)
\bibitem{melia} D.R. Ballantyne, et al., {\it Astrophys. J.} {\bf 657}, L13 (2007)
\bibitem{wolfendale} A.D. Erlykin, A. Wolfendale, {\it J. Phys. G: Nucl. Part. Phys.} {\bf 34}, 1813 (2007)
\bibitem{apostolos} S. Dimitrakoudis, et al., {\it Astropart. Phys.} {\bf 31}, 13 (2009)
\bibitem{3D} M. Tsuboi, et al., {\it Astrophys. J. Suppl.} {\bf 120}, 1 (1999); M.J. Reid, et al.,~{\it Astrophys. J.} {\bf 705}, 1548 (2009); G. Ponti, et al., {\it Astrophys. J.} {\bf 714}, 732 (2010)
\bibitem{hess} F.A. Aharonian, et al., {\it Astron. Astrophys.} {\bf 481}, 401 (2008)
\bibitem{fermi} A.A. Abdo, et al., {\it Astrophys. J.} {\bf 718}, 348 (2010)
\bibitem{agile} A. Giuliani, et al., {\it Astron. Astrophys.} {\bf 516}, L11 (2010)
\bibitem{sf2a} S. Gabici, et al., to appear in the Proc. of SF2A-2010 -- arXiv:1009.5291
\bibitem{fujita} Y. Fujita, et al., {\it Astrophys. J.} {\bf 707}, L179 (2009)
\bibitem{li} H. Li, Y. Chen, {\it Mon.Not. R. Astron. Soc.}, in press - arXiv:1009.0894
\bibitem{ohira} Y. Ohira, et al., {\it Mon. Not. R. Astron. Soc.}, in press - arXiv:1007.4869
\bibitem{diego} D.F. Torres, et al., {\it Mon. Not. R. Astron. Soc.} {\bf 387}, L59 (2008)
\bibitem{IC443GeV} A.A. Abdo, et al., {\it Astrophys. J.} {\bf 712}, 459 (2010)
\bibitem{IC443TeV} J. Albert, et al., {\it Astrophys. J.} {\bf 664}, L87 (2007)
\bibitem{yas} Y. Uchiyama, et al., arXiv:1008.1840
\bibitem{various} F.A. Aharonian, et al., {\it Astron. Astrophys.} {\bf 483}, 509 (2008); {\it ibid.}, {\bf 490}, 685 (2008); A. Fiasson, et al. 2009, Proc. XLIV Rencontres de Moriond, p. 81 
\bibitem{slane} D. Castro, P. Slane, {\it Astrophys. J.} {\bf 717}, 372 (2010)
\bibitem{cygnus} A.A. Abdo, et al., {\it Astrophys. J.} {\bf 688}, 1078 (2008)
\end{thebibliography}
\end{document}